\documentclass[useAMS,usenatbib]{mn2e}

\usepackage[psamsfonts]{amssymb}
\usepackage[dvips]{graphicx}
\usepackage{amsmath,alltt}
\usepackage{multirow}
\usepackage{rotating}
\usepackage{lscape}

\title[Illustrating chaos]{Illustrating chaos:  A schematic discretization of the general three-body problem in Newtonian gravity}
\author[Leigh N. W. C., Wegsman S.]{Nathan W. C. Leigh$^{1}$, Shalma Wegsman$^{1,2}$
\thanks{E-mail: nleigh@amnh.org (NWCL)}\\
$^{1}$Department of Astrophysics, American Museum of Natural History, Central Park West and 79th Street, New York, NY 10024 \\
$^{2}$Department of Astronomy and Astrophysics, University of Chicago, 5640 S. Ellis Ave., Chicago, IL 60637}
\begin{document}

\pagerange{\pageref{firstpage}--\pageref{lastpage}} \pubyear{2016}

\maketitle

\label{firstpage}

\begin{abstract}
We present a formalism for constructing schematic diagrams to depict chaotic three- body interactions in Newtonian gravity. This is done by decomposing each interaction in to a series of discrete transformations in energy- and angular momentum-space. Each time a transformation is applied, the system changes state as the particles re-distribute their energy and angular momenta. These diagrams have the virtue of containing all of the quantitative information needed to fully characterize most bound or unbound interactions through time and space, including the total duration of the interaction, the initial and final stable states in addition to every intervening temporary meta-stable state. As shown via an illustrative example for the bound case, prolonged excursions of one of the particles, which by far dominates the computational cost of the simulations, are reduced to a single discrete transformation in energy- and angular momentum-space, thereby potentially mitigating any computational expense. We further generalize our formalism to sequences of (unbound) three-body interactions, as occur in dense stellar environments during binary hardening. Finally, we provide a method for dynamically evolving entire populations of binaries via three-body scattering interactions, using a purely analytic formalism. In principle, the techniques presented here are adaptable to other three-body problems that conserve energy and angular momentum.
 
\end{abstract}


\section{Introduction} \label{intro}

After hundreds of years of study \citep[e.g.][]{newton1686}, the chaotic three-body problem remains pertinent to the field of stellar dynamics, as it relates to our understanding of interactions within globular \citep[e.g.,][]{heggie75,leigh12,leigh13b,antonini16}, open \citep[e.g.,][]{leonard89,hurley05,leigh11,geller13,leigh13a,mapelli14,geller15,leigh16} and even nuclear \citep[e.g.][]{davies98,merritt13} star clusters.  The problem remains analytically unsolved.  Historically, any solutions involved approximate methods or the application of constraining assumptions used to simplify the problem.  The best known such example of this approach is the restricted three-body problem, as applied to the Earth-Sun-Moon system.  But modern computer technologies have come a long way and, in so doing, sparked a new wave of intensive study of such systems over the last few decades.

Most simulations for the chaotic three-body problem run very quickly on modern computers.  Every now and then, however, you will encounter one simulation that gets hung up, and takes many orders of magnitude more computer time to run to completion.  Thus, upon performing a suite of numerical experiments of single-binary scattering, for example, it is a handful of individual runs that can consume the vast majority of the total computational cost.  For small tidal tolerance parameters (see Section~\ref{exp} for more details), this is due to prolonged excursions, where one of the stars ends up on a near-zero energy orbit.  In fact, some authors have argued that the mean disruption time should formally be infinite for resonant three-body interactions \citep[e.g.][]{valtonen75,valtonen06,shevchenko10,orlov10,leigh16}; if an infinite number of simulations could be performed, the probability of obtaining one or two simulations with near infinite excursion time events becomes finite. These individual events can be bottlenecks in studies of the chaotic three-body problem, and compound to become even worse when additional particles are included in to the mix.


Chaotic three-body systems interacting resonantly display a characteristic evolution in time that begins when one particle is sent on a temporary excursion while remaining bound to the system.  Upon returning to periastron, the particle initiates another close triple encounter, interacts strongly with the other two particles, and the process repeats \citep[e.g.][]{agekyan67,anosova94}.  To refer to events where the ejected particle is sent on a bound excursion we will use the term "ejection".  Conversely, to refer to events where the particle becomes unbound from the system and does not return, we will use the term "escape".  The end result of this chaotic dance is always an escape event where one of the bodies is ejected with a sufficiently high velocity to become unbound, and escapes to spatial infinity.  

The chaotic progression or evolution described above can be simplified or reduced to a single discrete transformation in energy and angular momentum space.  The initial and final states of the system are qualitatively the same, since the unbound system can be decomposed in to a single star and a binary.  Once the initial relative energies and angular momenta of the single star and the binary are decided (via the input parameters that define a given scattering experiment), a transformation is applied.  This changes the relative energies and angular momenta such that they are distributed differently among the particles in the initial and final states.  By neglecting the intermediary chaos that occurs in the bound state, statistical ensembles of outcomes can be considered directly, rather than relying on specific choices of scattering parameters.  

In this paper, we present a schematic formalism to discretize the general three-body problem in Newtonian gravity, in both the bound and unbound states. This is done in such a way that, for most cases of interest, our diagrams fully and quantitatively characterize the time evolution of a given chaotic single-binary interaction. We use the numerical scattering code \texttt{FEWBODY} to simulate a series of binary-single encounters involving identical point particles in Section~\ref{method}. We choose one particularly interesting example, that features a very prolonged excursion of one of the stars, and generate our schematic diagrams to depict the time evolution of this interaction. We further generalize our formalism to sequences of (unbound) three-body interactions, as occur in dense stellar environments during binary hardening, and even describe a purely analytic method for dynamically evolving entire populations of binaries via three-body scattering interactions. In Section~\ref{discussion}, we discuss the utility of our schematic formalism, address possible shortcomings and applications of our method and explain how our formalism might eventually replace the need for computer simulations of single-binary scatterings. We summarize in Section~\ref{summary}.
 
\section{Method} \label{method}

In this section, we introduce our formalism for constructing schematic diagrams of the general three-body problem, adapted from \citet{leigh12} and \citet{leigh16}.  We further present the numerical scattering experiments used to study the time evolution of the chaotic three-body problem, from which we construct an illustrative example of our schematic diagram formalism.  

\subsection{Constructing the schematic diagrams} \label{construct}

In this section, we begin by adapting the schematic diagrams first presented in \citet{leigh12} to depict individual snapshots of the system in time for application to the general three-body problem, including arbitrarily long excursions of one of the particles, followed by a description of our method for discretizing the interaction. 

First, the total energy of the three-body system can be written as:
\begin{equation}
\label{eqn:energy}
E = \sum_{\rm i=1}^{\rm N=3} E_{\rm i}' = \sum_{\rm i=1}^{\rm N=3} \Big( \frac{1}{2}m_{\rm i}v_{\rm i}^2 - \frac{1}{2}\sum_{\rm j=1, j \ne i}^{N=3} \frac{Gm_{\rm i}m_{\rm j}}{r_{\rm ij}} \Big),
\end{equation}
where $m_{\rm i}$ is the mass of the $i$-th particle, $v_{\rm i}$ is its velocity relative to the system centre of mass, $r_{\rm i}$ is the magnitude of its distance from the system centre of mass and $r_{\rm ij} = |\bar{r}_{\rm i} - \bar{r}_{\rm j}|$.  Re-writing Equation~\ref{eqn:energy}, we get:
\begin{equation}
\label{eqn:energy2}
E - E_{\rm 0} = \sum_{\rm i=1}^{\rm N=3} \Big(E_{\rm i}' - E_{\rm 0}/3 \Big) = \sum_{\rm i=1}^{\rm N=3} E_{\rm i},
\end{equation}
where $E_{\rm 0}$ is a negative constant with units of energy.  The addition of the constant $E_{\rm 0}$ serves only to shift the zero-point of the total encounter energy. This is needed to avoid the appearance of negative angles in our formalism.  We adopt this approach for the bound case in Section~\ref{bound}.  In Section~\ref{one} we consider a series of unbound three-body scattering interactions, and explore a different procedure for handling negative angles.  Here, negative angles instead correspond to unbound particles.  We will return to this in Section~\ref{one}.

Next, we point out that the sum of the angles of any polygon always add to a constant.  For example, the sum of the angles of a triangle sum to 180 degrees.  Similarly, the energy contained in each particle $E_{\rm i}$ must always add up to equal the (shifted) total encounter energy $E - E_{\rm 0}$.  Hence, we can write:
\begin{equation}
\label{eqn:angles}
\frac{E - E_{\rm 0}}{180^{\circ}} = \frac{E_{\rm i}}{\theta_{\rm i}},
\end{equation}
where $\theta_{\rm i}$ denotes the angle of a triangle corresponding to particle $i$ with total energy $E_{\rm i}$, as given by Equation~\ref{eqn:energy2}.

For each triangle, we fix the length of the side of the triangle oriented along the x-axis (i.e., parallel to the direction of increasing time) to be equal to unity. This ensures that, in energy-space, the symmetry of each triangle is positively correlated with the duration of the corresponding excursion event. More equilateral triangles correspond to longer excursion events (see Section~\ref{results} for more details).

A similar exercise can be performed to generate triangles for each component of the total angular momentum, namely $L_{\rm x}$, $L_{\rm y}$ and $L_{\rm z}$.  Using the x-component of the total angular momentum as an example we can write:
\begin{equation}
\label{eqn:angmom}
L_{\rm x} - L_{\rm x,0} = \sum_{\rm i=1}^{\rm N=3} \Big(L_{\rm x,i}' - L_{\rm x,0}/3 \Big) = \sum_{\rm i=1}^{\rm N=3} L_{\rm x,i},
\end{equation}
and hence:
\begin{equation}
\label{eqn:angles2}
\frac{L_{\rm x} - L_{\rm x,0}}{180^{\circ}} = \frac{L_{\rm x,i}}{\theta_{\rm x,i}},
\end{equation}
where $\theta_{\rm x,i}$ denotes the angle of a triangle corresponding to particle $i$ with an x-component of the total angular momentum $L_{\rm x,i}$, as given by Equation~\ref{eqn:angmom}.

Next, we describe our formalism for discretizing the time evolution of the interaction. As described in Section~\ref{intro}, the evolving three-body system can usually be decomposed in to a temporary binary, with the other star acting as a temporary single star on an excursion. This single-binary pair forms a closed (bound) temporary orbit. In order to formally define this criterion, we output from the simulations described in the subsequent section the critical time-steps at which the radial component of the centre of mass velocity of the temporary binary changes sign from positive to negative. This always coincides with the time-step at which the radial component of the centre of mass velocity of the temporary single star also changes sign. Every time this transition occurs, we record the energies of the particles and all three components of their total angular momentum. We emphasize that, since there are three particles, in this scenario there will always be one particle that is furthest from the system centre of mass, and is closest to the $E \sim 0$ boundary.

Using the formalism described above for generating triangles that quantitatively describe the temporary system in energy- and angular momentum-space, we then construct four triangles for each time-step corresponding to this change in sign of the radial velocities, one for energy and three for the total angular momentum (one for each component). The angle corresponding to each vertex of a given triangle is directly proportional to the energy or angular momentum content of the indicated particle, as described above. The angle corresponding to the top vertex is defined to always correspond to the temporary single star. The triangles are arranged sequentially from left to right, to indicate the direction of increasing time. The number of triangles is one greater than the number of applied transformations, to include the initial state of the system. This number indicates the total number of temporary excursion events, and clearly reveals any very long-lived computationally-expensive excursions (via the relative symmetry of the triangles; see above and Section~\ref{results}).

\subsection{Numerical scattering experiments} \label{exp}

We calculate the outcomes of a series of single-binary (1+2) encounters using the \texttt{FEWBODY} numerical 
scattering code\footnote{For the source code, see http://fewbody.sourceforge.net.}.  The code integrates the usual 
$N$-body equations in configuration- (i.e., position-) space in order to advance the system forward in time, using the 
eighth-order Runge-Kutta Prince-Dormand integration method with ninth-order error estimate and adaptive time-step.  
For more details about the \texttt{FEWBODY} code, we refer the reader to \citet{fregeau04}.  

\begin{figure}
\begin{center}
\includegraphics[width=\columnwidth]{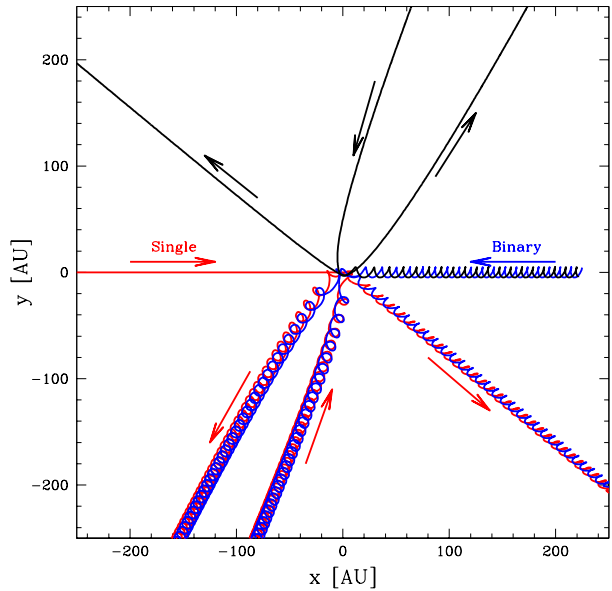}
\end{center}
\caption[Schematic diagram for the time evolution of a single-binary encounter in position-space, projected in to two-dimensions]{The time evolution of an example single-binary simulation in position-space, projected in to two-dimensions.  The unit of distance is 1 AU on both axes.  Notice the prolonged excursion of one of the particles, which dominates the total integration time of the simulation.  
\label{fig:fig1}}
\end{figure}

All objects are point particles with masses of 1 M$_{\odot}$.  The initial binary has $a_{\rm i} =$ 1 AU, and 
eccentricity $e_{\rm i} =$ 0.  We fix the impact parameter at $b =$ 0 for all simulations.  
The angles defining the initial relative configurations of the binary orbital plane and phases relative to the velocity vector of the incoming single star are 
chosen at random.  We perform 1000 numerical scattering experiments to find a suitable illustrative example, including a prolonged excursion of one of the particles.  This is illustrated in Figure~\ref{fig:fig1}.

To decide when an encounter is complete, we adopt the same criteria as \citet{fregeau04}.  To first order, we define this as the time when the separately bound hierarchies that constitute the system are no longer interacting with each other, or evolving internally.  Formally, the integration is terminated when the top-level hierarchies have positive relative 
velocity and the corresponding top-level $N$-body system has positive total energy.  
We set the tidal tolerance parameter to $\delta =$ 10$^{-7}$ for all simulations, which has been previously shown to be sufficient for the virial ratios (defined as the ratio between the total system kinetic energy and the total potential energy) considered here (see \citet{geller15} for more details).

\section{Results} \label{results}

In this section, we generate and present our schematic diagrams for the discretization of the general three-body problem. We begin with a single three-body interaction that remains bound and in a resonant state for several system crossing times, and apply our method to the illustrative example shown in Figure~\ref{fig:fig1}. We then present a modified version of our formalism to treat sequences of unbound single-binary scattering interactions, as occur during binary hardening in dense stellar environments, and show how (in some cases) our schematic diagrams can be generated purely analytically without any numerical simulations. Using this basic approach, we develop an analytic formalism for dynamically evolving entire populations of binaries in dense stellar environments due to three-body scattering interactions.

\subsection{Bound case} \label{bound}

\begin{figure}
\begin{center}                                                                                                                                                           
\includegraphics[width=\columnwidth]{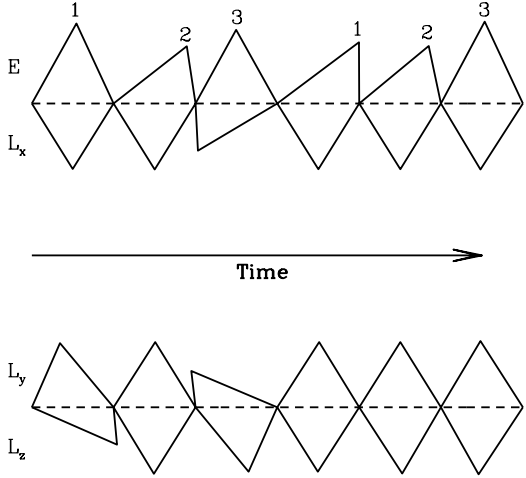}
\end{center}
\caption[Schematic diagrams for the time evolution of the single-binary encounter depicted in Figure~\ref{fig:fig1}]{The time evolution of the example single-binary simulation shown in Figure~\ref{fig:fig1} in both energy- and angular momentum-space, as described in the text. Four separate triangle series are shown, one for energy (top) and three for angular momentum (bottom), with one series for each of the three components. The $L_{\rm x}$ and $L_{\rm z}$ diagrams have been reflected about the horizontal dashed line. At the vertex of the apex of each triangle opposing the horizontal axis, we indicate which particle corresponds to the temporary single star, using the same choice of labeling as in Figure~\ref{fig:fig1}. Left to right denotes the direction of increasing time. We set $E_{\rm 0} = 6.6 \times 10^{38}$ J, $L_{\rm x0} = 4.0 \times 10^{48}$ kg m$^2$ s$^{-1}$, $L_{\rm y0} = 4.0 \times 10^{48}$ kg m$^2$ s$^{-1}$ and $L_{\rm z0} = 2.0 \times 10^{48}$ kg m$^2$ s$^{-1}$ for the zero-points, or equivalently $E_{\rm 0} = 10|E|$, $L_{\rm x0} = 2300L_{\rm x}$, $L_{\rm y0} = 510L_{\rm y}$ and $L_{\rm z0} = 1300L_{\rm z}$.
\label{fig:fig2}}
\end{figure}

Schematic diagrams for the single-binary interaction depicted in Figure~\ref{fig:fig1} are shown in Figure~\ref{fig:fig2}.  Four separate triangle series are shown, one for energy-space (top) and three for angular momentum (bottom) with one series for each of the three components. The $L_{\rm x}$ and $L_{\rm z}$ diagrams have been reflected about the horizontal dashed line. For each triangle, we fix the length of the side of the triangle oriented along the x-axis (i.e., parallel to the direction of increasing time). Combined with the imposed offset in energy $E_{\rm 0}$, this ensures that equilateral configurations of our triangles in energy-space are positively correlated with the duration of the corresponding excursion event. The initial and final states correspond to, respectively, the first and last triangles. Hence, the progression of time is from left to right.

The diagrams shown in Figure~\ref{fig:fig2} fully and quantitatively describe the time evolution of the example single-binary interaction depicted in Figure~\ref{fig:fig1}.  In principle, using these diagrams, the time evolution of the interaction in position- and velocity-space can be re-constructed.  This is because the energy and angular momentum of each particle is quantitatively encoded in the diagrams at each discrete exchange event via Equations~\ref{eqn:angles} and~\ref{eqn:angles2} combined with Equations~\ref{eqn:energy}, ~\ref{eqn:energy2} and~\ref{eqn:angmom}, and the duration of each corresponding excursion is quantitatively encoded by the the shape of each triangle (see below). \textit{Specifically, these diagrams encode 12 equations with 18 unknowns. The remaining 6 equations are found from the symmetry properties of the initial, final and temporary binaries (i.e., given the provided information in our diagrams, only the positions and velocities of one of the binary components are needed to compute those of the other).}  This is done as follows.  Consider a binary with component masses $m_{\rm 1}$ and $m_{\rm 2}$, oriented in the xy-plane.  We transform our coordinate system such that it is stationary in the centre-of-mass frame of reference of the binary.  Then, if the (instantaneous) positions and velocities of the component with mass $m_{\rm 1}$ are $(x_{\rm 1},y_{\rm 1},z_{\rm 1})$ and $(v_{\rm x,1},v_{\rm y,1},v_{\rm z,1})$, then the positions and velocities of the second component are:
\begin{eqnarray}
\label{eqn:comp2}
x_{\rm 2} = -\frac{m_{\rm 1}}{m_{\rm 2}}x_{\rm 1} \\
y_{\rm 2} = -\frac{m_{\rm 1}}{m_{\rm 2}}y_{\rm 1} \\
z_{\rm 2} = z_{\rm 1} = 0 \\
v_{\rm x,2} = -\frac{m_{\rm 1}}{m_{\rm 2}}v_{\rm x,1} \\
v_{\rm y,2} = -\frac{m_{\rm 1}}{m_{\rm 2}}v_{\rm y,1} \\
v_{\rm z,2} = v_{\rm z,1} = 0
\end{eqnarray}
This follows from the definition of the system centre of mass, which defines the equation:
\begin{equation}
\label{eqn:cofm}
m_{\rm 1}\vec{R_{\rm 1}} + m_{\rm 2}\vec{R_{\rm 2}} = 0,
\end{equation}
where $\vec{R_{\rm 1}}$ and $\vec{R_{\rm 2}}$ are the radius vectors of the bodies with respect to the binary centre of mass, and:
\begin{equation}
\label{eqn:radius}
|\vec{R_{\rm i}}| = \sqrt{x_{\rm i}^2 + y_{\rm i}^2 + z_{\rm i}^2},
\end{equation}
and $i =$ 1 or 2.

Now, consider the very prolonged excursion event indicated in Figure~\ref{fig:fig1}. This corresponds to the third triangle (from the left) in Figure~\ref{fig:fig2}. This is very close to a completely equilateral triangle in energy-space, to within fractions of a degree, whereas triangles corresponding to shorter excursion events are much less symmetric in energy-space. This behaviour is also evident by the corresponding triangles in $L_{\rm x}$-, $L_{\rm y}$- and $L_{\rm z}$-space, in which one particle clearly dominates the angular momentum budget during the excursion. We define the following parameter:
\begin{equation}
\label{eqn:epsilon} 
\epsilon = \sum_{\rm i=1}^{3} |\theta_{\rm i} - \frac{\pi}{3}|.
\end{equation}
The longest excursion events correspond to $\epsilon \sim 0$, whereas shorter excursion events satisfy $\epsilon > 0$. The reason for $\epsilon \sim 0$ corresponding to the longest excursions is that, here, the temporary single star is nearly unbound such that its energy is very close to zero. Meanwhile, the temporary binary is approximately isolated, such that it (approximately) obeys the virial theorem. Hence, twice the total kinetic energy is roughly equal to the absolute value of the total potential energy in the temporary binary. This translates into $E' \sim 0$ in Equation~\ref{eqn:energy}, and hence all $\theta_{\rm i}$ are approximately equal to ${\pi}/3$ via Equations~\ref{eqn:energy2} and~\ref{eqn:angles}.

\subsection{Unbound case: one binary} \label{one}

In this section, we briefly reproduce the analytic formalism for unbound scattering presented in \citet{valtonen06}, in particular the key formulae and distribution functions to be used and discussed in the subsequent sections.  These are presented here via an example application of our schematic formalism to the unbound case.  In dense stellar environments, as occur in the cores of star clusters, a single binary star can undergo repeated interactions with many different single stars. This sequence of single-binary scatterings typically acts to harden the binary (i.e., reduce its orbital separation). Our schematic diagrams can be adapted to quantitatively describe this process.

The required distribution functions are already available in analytic form for the initial and final unbound states (i.e., the first and last triangles) \citep{valtonen06}. This is thanks to pioneering efforts by \citet{monaghan76a} and \citet{monaghan76b} who re-worked the expression for the density of configuration states per unit energy in order to derive the distributions of binary orbital energies, eccentricities and escaper velocities in the low- and high-angular momentum limits. This was later parameterized to be applicable to any total angular momentum \citep{valtonen06}. Thus, we can already generate the initial and final states from these distribution functions.

One way to do this is as follows. We write for the total encounter energy of a given single-binary interaction:
\begin{equation}
\label{eqn:tot4}
E = E_{\rm s} + E_{\rm B},
\end{equation}
where $E_{\rm B}$ is the binary orbital energy and $E_{\rm s} = \frac{1}{2}m_{\rm s}v_{\rm s}^2$ is the kinetic energy of the single star in the centre of mass frame of the binary. From Equation 7.19 in \citet{valtonen06}, the distribution of escaper velocities is:
\begin{equation}
\label{eqn:vdist}
f(v_{\rm s})dv_{\rm s} = \frac{(n-1)|E|^{n-1}(m_{\rm s}M/m_{\rm B}))v_{\rm s}dv_{\rm s}}{(|E| + \frac{1}{2}(m_{\rm s}M/m_{\rm B})v_{\rm s}^2)^{n}},
\end{equation}
where $m_{\rm s}$ is the mass of the single star, $m_{\rm B}$ is the mass of the binary and $M = m_{\rm s} + m_{\rm B}$ is the total system mass. From Equation 7.26, we have for the final distribution of binary orbital energies:
\begin{equation}
\label{eqn:distEb2}
f(|E_{\rm B}|){d}|E_{\rm B}| = (n-1)|E|^{n-1}|E_{\rm B}|^{-n}{d}|E_{\rm B}|.
\end{equation}
Equations~\ref{eqn:vdist} and~\ref{eqn:distEb2} can be sampled from directly\footnote{Although, in order to conserve energy and momentum, it is easiest to sample from only one of them.  In our example case, we sample from the velocity distribution in Equation~\ref{eqn:vdist}, and use this to calculate a corresponding binary orbital energy using conservation of energy.}, and hence final states constructed from the initial state. The parameter $n$ corrects each distribution for the total angular momentum, via the simple equation:
\begin{equation}
\label{eqn:n}
n = 3 + 18L^2,
\end{equation}
and here $L$ is a normalized version of the total encounter angular momentum $\vec{L}$. Equation~\ref{eqn:n} has been calibrated from numerical scattering experiments of single-binary interactions \citep{valtonen06}.

Now, to construct the schematic triangles, we can set:
\begin{equation}
\label{eqn:thetaB}
\frac{E}{180^{\circ}} = \frac{E_{\rm s}}{\theta_{\rm s}} = \frac{E_{\rm B}}{\theta_{\rm B}},
\end{equation}
where $\theta_{\rm s}$ is the angle corresponding to the single star, which is found at the apex of the triangle, and $\theta_{\rm B} = \theta_{\rm B1} + \theta_{\rm B2}$ are the angles for each binary component, at the base of the triangle. We assume $\theta_{\rm B1} = \theta_{\rm B2}$ for this example, to keep things simple. If $E_{\rm s}$ exceeds the critical energy for escape, which is the situation for the unbound case, then $\theta_{\rm s}$ is negative. Hence, to correct for this, we set $\theta_{\rm s}' = -\theta_{\rm s}$ along with $\theta_{\rm B}' = 360^{\circ} - \theta_{\rm B}$.  We emphasize that, unlike in the bound case in Sections~\ref{construct} and~\ref{bound}, we allow for negative angles in this section for the unbound case, to treat unbound escaping stars appropriately (given that their total energy is positive, and we began by assigning positive angles to negative energies).

\begin{figure}
\begin{center}                                                                                                                                                           
\includegraphics[width=\columnwidth]{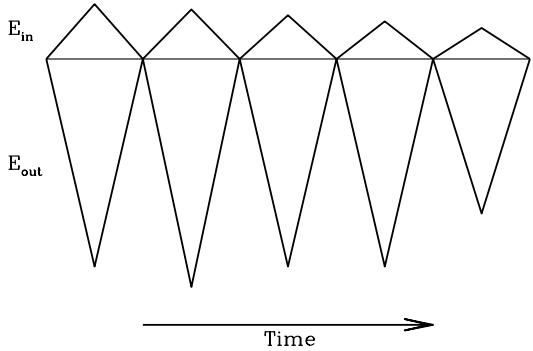}
\end{center}
\caption[Schematic diagrams for the unbound case in energy-space]{Schematic diagrams for the unbound case in energy-space. We show five successive binary hardening events due to single-binary scattering. The initial and final states for each scattering event are shown, respectively, at the top and bottom. The incoming and outgoing single stars always correspond to the vertex at the apex (i.e., furthest from the horizontal axis) of a given triangle. Left to right denotes the direction of increasing time.
\label{fig:fig3}}
\end{figure}

An example of our schematic diagrams for the unbound case is shown in Figure~\ref{fig:fig3}, which depicts five successive binary hardening events from single-binary scattering. All particles are assumed to have masses 1 M$_{\odot}$. The initial binary has a semi-major axis of 5 AU. We assume a velocity dispersion of 5 km s$^{-1}$ for the surrounding stellar environment which provides the single stars for each scattering interaction. Hence, we take $v_{\rm inf} =$ 5 km s$^{-1}$ as the initial relative velocity at infinity for every single-binary interaction. We further assume that, in every final state, the escaping single star is ejected with a velocity equal to the peak of the velocity distribution in Equation~\ref{eqn:vdist}, or:
\begin{equation}
\label{eqn:vpeak}
v_{\rm s,peak} = \alpha\sqrt{|E|\frac{M - m_{\rm s}}{m_{\rm s}M}}.
\end{equation}
We set $\alpha =$ 0.5, which corresponds to typical intermediate values of the angular momentum $L$, and hence to isotropic scattering.  We emphasize that this choice of adopting the peak velocity for the escaper (instead of drawing at random from the escaper velocity distribution in Equation~\ref{eqn:vdist}) is arbitrary, and is adopted only for simplicity in this illustrative example.

A similar formalism can potentially be applied in angular momentum-space, albeit this should be done with caution.  In particular, the exact methodology for this depends on exactly which distributions have been derived, and ultimately whether or not they are joint or marginalized distributions.  We have for the total angular momentum:
\begin{equation}
\label{eqn:angtot}
\vec{L} = \vec{L}_{\rm s} + \vec{L}_{\rm B}.
\end{equation}
From this, an analogous procedure can (in principle) be applied to define equivalent diagrams in angular momentum-space as done above in energy-space, for all three components of the angular momentum.

We emphasize that the formalism presented here for characterizing the time evolution of a given binary star system as it experiences scattering interactions with single stars in a dense stellar environment is purely analytic. Consequently, any potential issues related to computational expense are entirely mitigated, whereas this is never really the case for either large-scale N-body simulations of star cluster evolution or suites of individual numerical scattering simulations.

\subsection{Unbound case: many binaries} \label{many}

Consider applying our formalism for the unbound case to an entire population of binary star systems. This is equivalent to dynamically evolving a population of binary stars in a live dense stellar environment. Using the analytic distribution functions from the previous section for the properties of the products of three-body scattering interactions, we can construct a purely analytic formalism to propagate an initial distribution of binary orbital parameters forward through time due to single-binary scattering interactions. This is done as follows.

We begin with a population of $N$ binaries, and assume a distribution of orbital separations $f(a)$, a distribution of mass ratios $f(q)$ and a primary mass distribution $f(m_{\rm 1})$. These distribution functions can be convolved to obtain the distribution of binary orbital energies $f_{\rm B}(E_{\rm B})$, using the chain rule:
\begin{equation}
\label{eqn:chain}
f_{\rm B}(E_{\rm B}) = \Big( \frac{dN}{da} \Big)\Big( \frac{dE_{\rm B}}{da} \Big)^{-1} + \Big( \frac{dN}{dq} \Big)\Big( \frac{dE_{\rm B}}{dq} \Big)^{-1} + \Big( \frac{dN}{dm_{\rm 1}} \Big)\Big( \frac{dE_{\rm B}}{dm_{\rm 1}} \Big)^{-1}
\end{equation}
The rate of change of $f_{\rm B}(E_{\rm B})$ is:
\begin{equation}
\label{eqn:rate}
\frac{df_{\rm B}(E_{\rm B})}{dt} = \frac{df_{\rm B}}{dE_{\rm B}}\Big( \frac{dE_{\rm B}}{dN} \Big)\Big( \frac{dN}{dt} \Big) = \frac{df_{\rm B}}{dE_{\rm B}}\Big( \frac{\Gamma(E_{\rm B})}{f(|E_{\rm B}|)} \Big),
\end{equation}
where $\Gamma(E_{\rm B})$ is the rate of single-binary interactions \citep{leigh11} and $f(|E_{\rm B}|)$ is the distribution of binary orbital energies given in Equation~\ref{eqn:distEb2}. The interaction rate $\Gamma$ depends on the binary orbital energy via the orbital separation and total binary mass, in addition to the properties of the host stellar environment (density, velocity dispersion, etc.).  

The angular momentum dependence enters via the assumed impact parameter distribution $f(b)$. For isotropic scattering in spherical star clusters, this typically takes the form $f(b)db = bdb$ and is related to the total angular momentum via $L = {\mu}v_{\rm inf}b$, where $\mu$ is the reduced mass of the initial single-binary system. Ultimately, these relations can be used to relate the assumed impact parameter distribution to a distribution of power-law indices $n$ in Equation~\ref{eqn:distEb2}, via Equation~\ref{eqn:n}.

The only requirement for the above analytic methodology to be accurate, is for the single-binary interactions to enter a long-lived "resonant" state, in which all particles are thoroughly mixed in phase space and the assumption of ergodicity is upheld. Fortunately, the cross-section for such a resonant encounter is straight-forward to calculate analytically \citep{hut83b}.

\section{Discussion} \label{discussion}

In this paper, we present a schematic discretization of the general three-body problem in Newtonian gravity. Our diagrams fully and quantitatively describe the time evolution of chaotic three-body interactions involving point particles for many, if not most, encounters. Hence, using our diagrams, the time evolution of most (bound or unbound) interactions can in principle be reconstructed in position- and velocity- space. In this section, we address the shortcomings and utility of our "chaos diagrams", discuss possible applications of our method and look to future work.

\subsection{The $E = 0$ boundary} \label{nearzero}

Particles can end up asymptotically approaching the $E = 0$ boundary, although the probability for such an event is vanishingly small in the limit $E \rightarrow 0$ for negative total encounter energies. The duration of these asymptotic excursions can be exceedingly long, and by far dominate the total computer run times for simulations.  Nearly all of the computational expense lies in these prolonged excursions.

Our schematic formalism reduces these prolonged excursion events to single discrete transformations in energy- and angular momentum-space. Thus, if numerical scattering simulations of three-body interactions were to be replaced with our chaos diagrams, this computational cost would be altogether mitigated.

\subsection{How to perform gravitational scattering experiments without a computer} \label{without}

For many, if not most, resonant three-body interactions, all of the quantitative information is encoded in our chaos diagrams.  Thus, in principle, these diagrams could remove the need to perform computer simulations, at least for some regions of the total relevant parameter space. It is just a matter of generating the distributions of particle energies and angular momenta at each discrete excursion event, without relying on computer simulations. This amounts to somehow calculating distribution functions for these quantities, from which we can sample at each discrete excursion event. This has already been done semi-analytically for the unbound case, which has provided simple analytic functions to work with (see Section~\ref{one} and Equation~\ref{eqn:distEb2}). No such analytic functions are known for the bound case. If such a function is derived, however, then in principle the formalism presented in this paper could completely replace numerical scattering simulations in those regions of parameter space where the assumption of ergodicity is upheld. For this to be the case, two additional criteria must also be satisfied:

\begin{itemize}

\item The approximation $E \sim E_{\rm s} + E_{\rm B}$ must remain valid throughout the entire (bound) interaction, such that each transformation in energy- and angular momentum-space corresponds to a single orbit of the (temporary) single star about the (temporary) binary. The state of each such temporary orbit is evaluated at apocentre.

\item An analytic formalism for mapping the system from one region of phase space to another between successive transformations (i.e., temporary ejection events) during the bound state.  For example, one way this could occur is if each transformation in energy- and angular momentum-space turns out to be independent of energy and angular momentum, and correspond to a random sampling of the corresponding distribution functions. That is, the different meta-stable states corresponding to each discrete event are independent, such that the probability of the system being in one region of parameter space does not affect the probability of it finding itself in another after a given transformation.

\end{itemize}

The first criterion is not always upheld, and it can occur that all three particles have comparable energy and angular momenta, such that the approximation $E \sim E_{\rm s} + E_{\rm B}$ temporarily breaks down during the bound state. Thus, currently, the formalism presented in this paper cannot do this accurately, since we ignore cross-terms in the energy equation that become more important during very short excursions.  However, as already stated, for many three-body interactions this assumption is strictly upheld for the entire duration of the encounter. In future work, we intend to study this interesting limit of the chaotic three-body problem in more detail. We intend to better define and quantify this transition, in order to understand how to include in our schematic formalism this more ambiguous phase in the evolution of the system.  

\begin{figure}
\begin{center}                                                                                                                                                           
\includegraphics[width=\columnwidth]{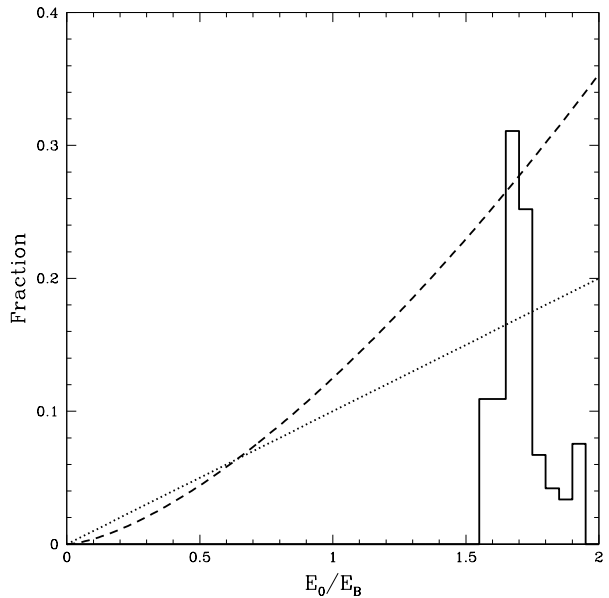}
\end{center}
\caption[The distribution of binary orbital energies for the bound state]{The distribution of binary orbital energies for the bound state, normalized by the total encounter energy as $z = |E|/|E_{\rm B}|$. We show the results of a single very prolonged three-body interaction, with 120 independent discrete transformations in energy- and angular momentum-space. We adopt a bin size of ${\Delta}z = 0.05$. The dotted ($n = 3$) and dashed ($n = 3.5$) lines show the binary orbital energy distribution function given by Equation~\ref{eqn:distEb2} for the unbound case in the low angular momentum limit, for two different values of the power-law index $n$.
\label{fig:fig4}}
\end{figure}

The second criterion can in principle be tested using numerical scattering experiments via the procedure described in Section~\ref{method}. In an effort to quantify and illustrate this, we identified one especially prolonged single-binary interaction from our suite of simulations. This yielded a total of 120 discrete transformation events (i.e., triangles in Figure~\ref{fig:fig2}), and provided the needed statistics to get an approximate idea of the distribution of binary orbital energies during the bound state of a given encounter, as shown in Figure~\ref{fig:fig4}. As is clear, the simulated distribution deviates from what is expected from a smooth continuation of Equation~\ref{eqn:distEb2} from the bound state to the unbound state, and resembles more of a skewed Gaussian. Unfortunately, the reason for this is not clear, and we do not know how sensitive it is to our adopted criterion for sampling the binary orbital energies.  As already stated, clearly defining an excursion event can become ambiguous during very short excursions.  The data shown in Figure~\ref{fig:fig4} naively suggest that the system remains close to its initial location in phase space throughout the entire resonant phase of the interaction.  More work will be needed to verify and better understand the basic results shown and discussed here. In the subsequent sections, we discuss two possible methods for better accomplishing the over-arching goal of quantifying the distributions of binary orbital energies and angular momenta during the bound state.

\subsubsection{Embracing the machine} \label{embrace}

Motivated by our results in this paper and especially Figure~\ref{fig:fig4}, one way to obtain distribution functions to describe the probabilities of each particle having a given combination of energy and angular momentum during a given excursion event is machine learning.  That is, using a machine learning algorithm to calculate these distribution functions directly from millions (and preferably billions) of three-body simulations, while implementing supervised learning techniques.  For example, an artificial neural network learning algorithm could be implemented, with the inter-connections between artificial neurons defined by repeated fine-graining of different volume elements of the total phase space.  This could be used to evaluate the degree to which these regions can (or cannot) be causally connected by a single discrete transformation, for each level or degree of fine-graining.  In other words, the machine could be trained to learn how the different meta-stable states corresponding to each discrete event are inter-related, and to what extent (if any) the probability of the system being in one region of parameter space affects the probability of it finding itself in another after a given transformation.  This offers an ideal method for testing our assumption of ergodicity during the bound state; that is, that the available phase space should be populated uniformly.  Thus, this method will not only reveal the underlying patterns characteristic of the time evolution of the system throughout phase space, but also the underlying physical mechanism(s) driving this evolution.  

With the necessary distribution functions in hand, an operator could potentially be generated that propagates the system forward through time, transforming one triangle into the next.  Consider the following transformation $\hat{T}$ applied to a three-dimensional vector $\vec{\theta} = [\theta_{\rm 1},\theta_{\rm 2},\theta_{\rm 3}]$: \\
\begin{equation}
\hat{T}\left[
\begin{array}{c}
 \theta_{\rm 1} \\
 \theta_{\rm 2} \\
 \theta_{\rm 3} 
\end{array} \right] =
\left[ 
\begin{array}{c}
 \theta'_{\rm 1} \\
 \theta'_{\rm 2} \\
 \theta'_{\rm 3}
\end{array} \right]   
\end{equation}
The primes denote the vector components in the new transformed state.  Using machine learning, an operator $\hat{T}$ could be generated for each of the four three-dimensional vectors relevant to our problem, namely $E$, $L_{\rm x}$, $L_{\rm y}$ and $L_{\rm z}$.  Each operator would be applied to the corresponding vector at each discrete excursion event, transforming it to a new state.  

Machine learning algorithms are an option potentially worth exploring more, since they offer one avenue toward achieving the over-arching goal of eventually being able to generate our chaos diagrams without any intervention from computers. But the irony is rich. Below, we explore another promising option for accomplishing our goal, which could potentially take us all the way beyond the machine.

\subsubsection{Beyond the machine} \label{beyond}

In Section~\ref{one}, we used analytic distribution functions from \citet{valtonen06} for the products of single-binary interactions to calculate the energies of the particles in the final state, and presented chaos diagrams for the initial and final states using a purely analytic approach. As already discussed, to complete a purely analytic approach, this leaves only the intermediate bound meta-stable states, in which the ejected single star eventually returns to interact with the binary at least once more. If this distribution can be derived, we can reduce the creation of the chaos diagrams presented in Figure~\ref{fig:fig2} to a purely analytic problem. We emphasize that this could entirely replace the need for computer simulations, for interactions that satisfy the criteria discussed in the preceding sections. This can already be done in energy-space (and angular momentum-space), as shown in Section~\ref{one} for the unbound states.

In a forthcoming paper (Stone \& Leigh 2018, in prep), we re-visit the original \citet{monaghan76a} formalism for calculating the density of configuration states per unit energy.  A more complete description of the outcome distributions can be derived by incorporating an angular momentum-dependence in to the integrand in the expression for the density of states. The hope is that this will provide the required distribution functions for the angular momenta of the final binary and the ejected single star, in addition to the corresponding distribution functions for both energy and angular momentum for the bound case.  These joint distributions can in turn be marginalized over to calculate specific distribution functions of interest.

We note that, if the assumption of ergodicity is upheld collectively for both the bound and unbound states, then the mean number of excursions for a collection of three-body interactions should be equal to the total volume in phase space available to the system (i.e. bound and unbound states) divided by the total volume corresponding to bound states alone. This potentially provides a means of calculating an expectation value for the number of excursion events for a given (resonant) three-body interaction.

The potential to bypass the need for computers in studying the chaotic three-body problem is tempting, to say the least. If the bound state distribution functions (i.e., energy, angular momentum, as well as the duration and number of excursions) can indeed be derived using a First Principles approach, then our entire formalism could become purely analytic. Numerical simulations would only be needed to better understand these key distribution functions, and verify their accuracy. This will be the focus of future work.  Finally, we point out that, although the diagrams presented here were primarily used for illustrative purposes throughout this paper, they have the potential to fully depict the full time evolution of individual three-body interactions, along with \textit{all} of the associated quantitative data needed to uniquely identify any one interaction and even reproduce it as a visualization of the time evolution of the three-body system in position-space (i.e., what is often done to generate a visual aid for these types of interactions).  We would argue, however, that this information can be conveyed to an audience using our diagrams much more quickly and efficiently than upon using a time-series visualization, such as a movie.  The underlying formalism of our method also provides an extremely efficient means of storing the data needed to fully and quantitatively reproduce a given scattering interaction, compared to other output files in common use, with the positions outputted at regular time-steps.

\section{Summary} \label{summary}

In this paper, we present a formalism for constructing schematic diagrams to depict chaotic three-body interactions in Newtonian gravity.  This is done by decomposing each interaction in to a series of discrete transformations in energy- and angular momentum-space.  With each transformation, the system changes state as the particles re-distribute their energy and angular momentum.  These diagrams have the virtue of containing all of the quantitative information needed to fully characterize many (if not most) bound or unbound interactions through time and space, including the total duration of the interaction, the initial and final stable states in addition to every intervening temporary meta-stable state. As shown via an illustrative example for the bound case, prolonged excursions of one of the particles by far dominates the computational cost of the simulations. In our formalism, these are reduced to a single discrete transformation in energy- and angular momentum-space, thereby potentially mitigating any computational expense. We further generalize our formalism to sequences of (unbound) three-body interactions, as occur in dense stellar environments during binary hardening. Finally, we provide a method for dynamically evolving entire populations of binaries via three-body scattering interactions, using a purely analytic formalism. In principle, the techniques presented here are adaptable to other three-body problems that conserves energy and angular momentum.

\section*{Acknowledgments}

We thank an anonymous reviewer for very helpful comments and suggestions that improved the quality of our work.  We further thank Johan Samsing and Nicholas Stone for an early read of our manuscript and critical feedback.  N.~W.~C.~L. acknowledges support from a Kalbfleisch Fellowship at the American Museum of Natural History and the Richard Gilder Graduate School, as well as support from National Science Foundation Award AST 11-09395. 


\bsp

\label{lastpage}

\end{document}